# Prediction of Li$_3$Fe$_8$B$_8$ compound with rapid one-dimensional ion diffusion channels

Shiya Chen[1#], Paul Oftedahl[2#], Zhen Zhang[3], Zepeng Wu[1], Junjie Jiang[1], Vladimir Antropov[4], Julia V. Zaikina,[2] Shunqing Wu[1], Kai-Ming Ho[3], Yang Sun[1*]

[1]*Department of Physics, Xiamen University, Xiamen 361005, China*
[2]*Department of Chemistry, Iowa State University, Ames, Iowa 50011, United States*
[3]*Department of Physics, Iowa State University, Ames, IA 50011, United States*
[4]*Ames National Laboratory, U.S. Department of Energy, Ames, IA 50011, United States*
(Dated: September 20, 2025)

Using a computational crystal structure search in the Li-Fe-B ternary system, we predict a stable phase of Li$_3$Fe$_8$B$_8$, featuring 1D channels that enable rapid Li-ion transport. *Ab initio* molecular dynamics simulations show that the Li-ion diffusion coefficient in Li$_3$Fe$_8$B$_8$ surpasses that of common electrode and conductive additive materials by several orders of magnitude. The high diffusion in Li$_3$Fe$_8$B$_8$ can be explained by the Frenkel–Kontorova model, which describes an incommensurate state between the Li diffusion chain and the periodic potential field caused by the FeB backbone structure. The favorable lithium-ion diffusivity and mechanical properties of Li$_3$Fe$_8$B$_8$ make it a promising conductive additive for battery materials. Its itinerant ferromagnetism also offers a platform for exploring correlated-electron magnetism and spin-dependent phenomena.

## I. Introduction

In the 1970s, Beyeler *et al.* found that conducting ions in potassium hollandite are restricted to linear channels [1], sparking widespread research into similar one-dimensional (1D) systems. Materials with 1D channels have driven significant advancements due to the high ion diffusion rates in such materials. For instance, single-walled zeolitic nanotubes with 1D channels exhibit much higher diffusion rates and surface areas than those of traditional two-dimensional and three-dimensional zeolites [2], increasing their effectiveness for adsorption and catalysis. The self-assembled 1D channel zeolite mordenite enables highly efficient carbon dioxide sieving [3]. The interesting properties of the 1D channel structure extend beyond materials science. Recent studies in Earth sciences suggest that the mantle's silica-water superstructure may contain 1D channels for superionic water transport [4], offering new insights into water cycling in the deep Earth. In biology, the potassium ion channels exhibit 1D ion conduction that enables efficient, directional potassium transport, which is vital for maintaining membrane potential and facilitating neural signaling [5].

Known for their excellent structural stability and rapid ion diffusion, 1D channel structures hold promising application potential in energy storage, such as lithium-ion batteries (LiBs). Notably, fluorides [6] and oxides [7] with 1D channels demonstrate strong intercalation and fast-charging capabilities, while copper-coordinated cellulose [8] and carbon black organic porous electrolytes [9] provide high ionic conductivity and prevent dendrite growth in solid-state batteries. Recently, boride materials have attracted great interest due to their diverse properties, such as exfoliation potential toward two-dimensional MBenes and promising LiB performance [10–14]. However, the bulk counterpart of MBenes, MAB (group-A metals and metalloids – transition metals – borides), remains challenging to synthesize because of the mismatched reacting temperatures between group-A metals and refractory boron. Novel synthesis routes have recently been opened to access the largely unexplored group-I/II MAB phases together with computational predictions [15–23]. Battery applications have been developed in some related bulk boride systems. For example, Ni$_{n+1}$ZnB$_n$ phases have been fabricated to exhibit microporous structures and applied for LiB anodes [24]. A compound reported as Li$_{1.2}$Ni$_{2.5}$B$_2$ compound with 1D Ni-B honeycomb channels has been shown to exhibit rapid lithium-ion diffusion and outstanding cycling stability as a LiB anode material [25]. Mg-Fe-B compounds with 1D channels have been predicted to exhibit large charge capacity and fast Mg-ion diffusion as Mg battery cathodes [26]. These discoveries highlight the promising capabilities of borides as innovative materials in battery technology. Even closer to the ternary system that is the focus of this work, the Li-Co-B system was the subject of a recent computational study where one thermodynamically stable and two

---

[#]Equal contribution.
[*]Email: yangsun@xmu.edu.cn

metastable ternary phases were predicted and evaluated for potential as LiB anode materials [27].

In this study, we performed crystal structure searches and *ab initio* calculations to explore the previously uncharted Li–Fe–B phase diagram. We searched for stable phases with 1D channel structures and analyzed the mechanisms that could lead to enhanced lithium-ion diffusion in these phases. This paper is organized as follows. In Sec. II, we outline the theoretical methods used for structure prediction and *ab initio* calculations, along with experimental procedures followed for the synthesis attempts. In Sec. III, we present a new phase diagram for the Li–Fe–B system, analyze the properties of Li$_3$Fe$_8$B$_8$ compounds, discuss the results of the experimental synthesis attempts and propose future pathways to realize the predicted compounds. Finally, we present our conclusions in Sec. IV.

## II. Methods
### A. Construction of Li-Fe-B phase diagram

The high-throughput chemical element substitutions are performed with ternary boride structures to search Li-Fe-B ternary phases. We extract all ternary borides, XYB (B=boron) from the Material Project database [28]. The substitution strategy is implemented through three steps. Firstly, phases containing F, Cl, Br, I, O, S, N, P, C, and H elements are excluded because of the distinct chemical environments of these elements from Li and Fe. Secondly, for structures containing one 3*d* transition metal, we substitute the transition metal with Fe and the other element with Li, resulting in 590 structures. After structural optimization, only 66 unique structures remained. Thirdly, for compositions lacking the 3*d* transition metals, we replaced X and Y with Li and Fe, respectively, and switched them, producing 1152 Li-Fe-B structures, from which only 112 unique types were identified. The application of these strategies yielded a dataset of 172 unique Li-Fe-B phases without any redundancy.

The substituted Li-Fe-B structures were optimized by spin-polarized density functional theory (DFT) calculations using the VASP code [29,30], which employs the projector augmented wave (PAW) method [31]. The exchange and correlation energy is treated with the spin-polarized generalized gradient approximation (GGA) and parameterized by the Perdew−Burke−Ernzerhof formula (PBE). [32] A plane-wave basis was used with a kinetic energy cutoff of 520 eV, and the convergence criterion for the total energy was set to $10^{-5}$ eV. Monkhorst−Pack's sampling scheme [33] was adopted for Brillouin-zone sampling with a ***k***-point grid of $2\pi \times 0.033$ Å$^{-1}$. The lattice and atomic coordinates are fully relaxed until the force on each atom is less than 0.01 eV/Å.

The formation energy $E_f$ of ternary Li$_x$Fe$_y$B$_z$ was calculated by

$$E_f = \frac{E(\text{Li}_x\text{Fe}_y\text{B}_z) - xE(\text{Li}) - yE(\text{Fe}) - zE(\text{B})}{x+y+z}, \quad (1)$$

where $E(\text{Li}_x\text{Fe}_y\text{B}_z)$ is the energy of Li$_x$Fe$_y$B$_z$. $E(\text{Li})$, $E(\text{Fe})$, and $E(\text{B})$ are the energy of Li, Fe, and B ground-state bulk phases, respectively. To characterize the energetic stability of Li$_x$Fe$_y$B$_z$, we calculated the formation energy differences with respect to any reference phases forming the Gibbs triangle on the convex hull (denoted as $E_{\text{hull}}$). Thus $E_{\text{hull}} > 0$ indicates the distance above the convex hull. If $E_{\text{hull}} = 0$, Li$_x$Fe$_y$B$_z$ is a new ground state and the convex hull was updated. The reference phases of the convex hull were obtained from the Materials Project [28]. To verify the dynamic stability of the thermodynamically promising compounds, phonon dispersion curves were computed using a supercell finite displacement method with PHONOPY code [34].

### B. *Ab initio* molecular dynamics

To calculate the diffusion coefficient, *ab initio* molecular dynamics (AIMD) simulations were carried out using the PAW PBE functional within VASP. The electronic temperature is set equal to the ionic temperature, employing the Mermin functional. A 304-atom $2 \times 2 \times 2$ supercell of the Li$_3$Fe$_8$B$_8$ phase discovered in this study was used for the simulations, with the Γ-point sampling the Brillouin zone. The simulations were performed in the NVT ensemble at 300 K. Each simulation runs for over 10 *ps* after thermal equilibration, with a time step of 1.5 *fs*.

The diffusion coefficient was calculated from the AIMD trajectories as [35]

$$D = \frac{1}{2bN} \lim_{t \to \infty} \frac{d}{dt} \left[ \sum_{i=1}^{N} \sum_{k=1}^{M} \left( \Delta \vec{r}_{i,k}(t) \right)^2 \right], \quad (2)$$

where $b$ is the dimensionality factor. $N$ is the number of mobile ions. $M$ is the total number of steps that the mobile ions traveled, and $\Delta \vec{r}$ is the displacement of the mobile ions. $D$ is obtained by a linear fitting of the time-dependent mean squared displacement (MSD).

### C. Experimental methods

For the experimental attempts to synthesize Li$_3$Fe$_8$B$_8$, powders of LiH (ThermoFisher, 99.4% metals basis), Fe (Alfa Aesar, 22 mesh, 99.998% metals basis), and B (ThermoFisher, amorphous & crystalline, 325 mesh, 98% metals basis) were weighed in a 4:8:8 molar ratio for a total mass of 0.3 grams per sample. All manipulations were performed under argon in a glovebox with < 1 ppm oxygen. The powders were placed in polystyrene grinding vials with a polystyrene ball and a slip-on cap and double-sealed inside polypropylene bags in argon atmosphere. The vials were brought out of the glovebox and ball-

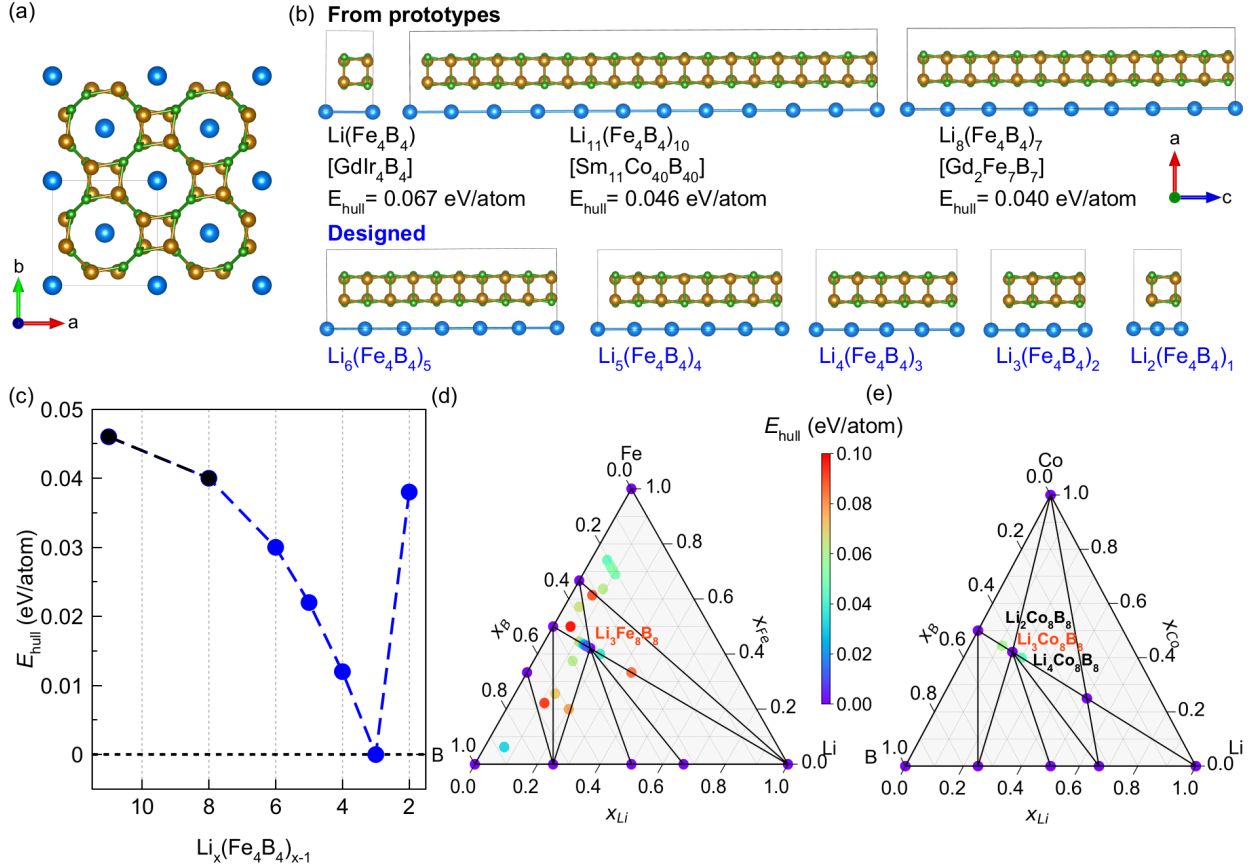

**FIG. 1** (a) Top view of $Li_x(FeB)_y$-type structures. Blue is Li. Brown is Fe. Green is B. (b) A series of structures with different numbers of Li and $(Fe_4B_4)$ units. The upper panel shows the systems from substituted structures prototype structures. The lower panel shows the designed $Li_x(FeB)_{x-1}$ phases. (c) The energy above the convex hull, $E_{hull}$, with respect to the Li content in $Li_x(Fe_4B_4)_{x-1}$. (d) Convex hull of the Li-Fe-B system including the new structures. (e) The updated Li-Co-B convex hull.

milled using a SPEX Sample Prep 8000 M MIXER/MILL for 18 minutes. The vials were then brought back into the glovebox and the powders were sealed inside Nb crucibles by arc-welding. Each crucible was then placed in a silica reactor equipped with a Swagelok one-way check valve to prevent the buildup of hydrogen pressure and evacuated to below $3.5 \times 10^{-5}$ bar. The reactor and crucible were placed in a resistance furnace (Thermo Scientific Thermolyne Type FD1500M) connected to a thermocontroller (Eurotherm 3216) and heated to different temperatures, dwelt for different durations followed by turning the furnace off and cooling a sample in the furnace: sample A (heating to 1023 K at 2.0 K/min, dwelling for 72 hours), sample B (heating to 1073 K at 1.6 K/min, dwelling for 24 hours), sample C (heating to 1173 K at 1.8 K/min, dwelling for 24 hours), and sample D (heating to 1273 K at 2.0 K/min, dwelling for 24 hours). The Nb crucibles were opened inside the glovebox to remove the product, and the powdered samples were ground gently with a mortar and pestle.

Powder X-ray diffraction (PXRD) was accomplished using a Rigaku MiniFlex600 powder diffractometer with Cu $K_\alpha$ radiation ($\lambda = 1.54051$ Å) and a Ni $K_\beta$ filter. Zero-background plate holders composed of silica or a silicon single crystal were used for data collection in ambient conditions, and detector settings were chosen to minimize fluorescence of iron.

### III. Results and discussions
#### A. The phase diagram of the Li−Fe−B system

We construct the convex hull phase diagram for the Li-Fe-B ternary system (see Supplementary Material Fig. S1 [36]) and compute the relative energy with respect to any reference phases forming the Gibbs triangle on the convex hull (denoted as $E_{hull}$). Although no stable ternary phase was found by the substitution of prototypical structures, we identify a class of low-energy metastable structures represented by $Li_x(FeB)_y$ on the tie line of Fe and LiB. These include $LiFe_4B_4$ substituted from the $GdIr_4B_4$ structure, $Li_{11}Fe_{40}B_{40}$ substituted from the $Sm_{11}Co_{40}B_{40}$ structure, and $Li_2Fe_7B_7$ substituted from the $Gd_2Fe_7B_7$ structure. In these structures, Fe and B atoms form backbones along the c-axis, with adjacent Fe-B bonds interconnected with each other. The large interstitial spaces in the framework create 1D channels along the c-axis, where Li atoms form a single chain, as shown in Fig. 1(a) and (b).

Inspection of these metastable structures suggests the repeating Fe-B backbone structure along the c-axis is composed of a $(Fe_4B_4)$ unit. The $Li_2Fe_7B_7$ phase is $Li_8(Fe_4B_4)_7$, and the $Li_{11}Fe_{40}B_{40}$ phase can be represented as $Li_{11}(Fe_4B_4)_{10}$. Compared to the energy of $Li_1(Fe_4B_4)_1$ ($E_{hull}$=0.067 eV/atom), $Li_{11}(Fe_4B_4)_{10}$ and $Li_8(Fe_4B_4)_7$ show lower energy to the convex hull, as 0.046 and 0.040 eV/atom, respectively. Thus, a mismatched Li and $Fe_4B_4$ population by one can be a way to generate a low-energy structure for the Li-Fe-B system. With this observation, we design a series of new structures, i.e., $Li_6(Fe_4B_4)_5$, $Li_5(Fe_4B_4)_4$, $Li_4(Fe_4B_4)_3$, $Li_3(Fe_4B_4)_2$, and $Li_2(Fe_4B_4)_1$, as shown in Fig. 1(b). The energy of these phases is shown in Fig. 1(c). The phase stability increases almost linearly with increasing Li content up to $Li_3(Fe_4B_4)_2$. At the $Li_3(Fe_4B_4)_2$ composition, a hull distance minimum is found.

As determined previously, no stable ternary phase was found in the Li-Fe-B convex hull diagram [28,37]. After recalculating the convex hull with new phases, the phase with composition $Li_3(Fe_4B_4)_2$ stands out as a thermodynamically stable phase as shown in Fig. 1(d). For simplicity, it is referred to as $Li_3Fe_8B_8$ hereafter. The crystallographic data is shown in Supplemental Material Table S1 [36]. We perform phonon calculations [34] to investigate the dynamical stability of the $Li_3Fe_8B_8$ phase. As shown in Fig. 2(a), the absence of imaginary frequencies in the phonon spectrum indicates that the structure is dynamically stable.

We studied several collinear magnetic orderings and the ground state of $Li_3Fe_8B_8$ appears to be ferromagnetic with a magnetic moment of 0.8 $\mu_B$/Fe (see Supplemental Material Fig. S2 and Table S2 [36]). Magnetic moments in AFM structures are much smaller (0.1-0.5$\mu_B$) and strongly depend on configuration. This indicates that magnetism in our system has a strong itinerant nature. Moreover, a significant magnetic short-range order can be expected [38], similar to the famous iron pnictides. In addition, such a relatively small moment for the Fe atom and rather good metallicity (see the electronic density of state in Fig. 2(b)) can lead to the appearance of notably spin fluctuations. Such fluctuations are not included in normal DFT and can strongly decrease the value of the static magnetic moment from above [39]. Thus, in the magnetic sense, the system represents an itinerant ferromagnet near the magnetic quantum critical point. The magnetic properties of such systems can be tuned by applied magnetic field, pressure or doping. This tunability can be leveraged in the manufacturing and recycling of corresponding lithium-ion batteries [40].

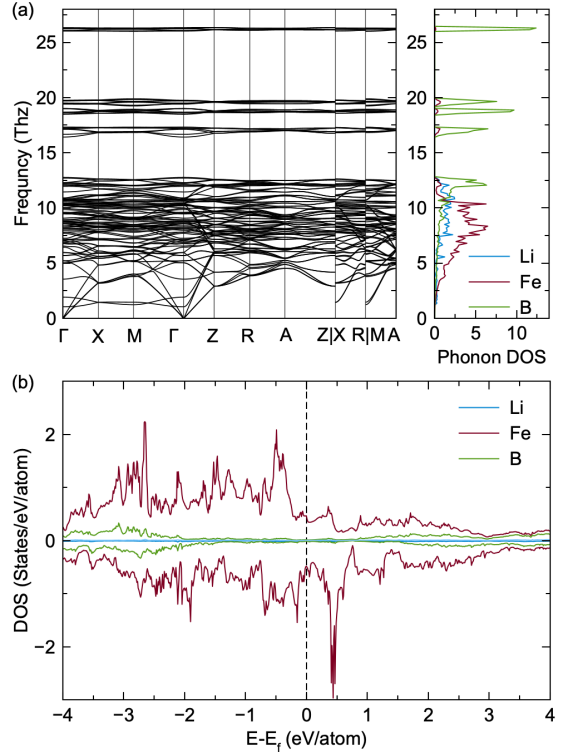

**Fig. 2** (a) Phonon band structure and density of state of $Li_3Fe_8B_8$ phase. (b) Electronic density of state of ferromagnetic $Li_3Fe_8B_8$.

In addition to the stable $Li_3Fe_8B_8$, the $Li_x(FeB)_{x-1}$ phase with $x$ ranging from 2 to 11 all shows $E_{hull}$ less than 0.05 eV/atom. Compounds with such low $E_{hull}$ suggest a high possibility of being synthesized in experiments, if the reaction barrier can be overcome [10,17,41,42]. Considering the similarity among transition metals, we replace Fe with Co to examine the stability of 3-8-8 motif in the Li-Co-B system. Starting from the $Li_{2-4}Fe_8B_8$ prototypes, direct replacement of Fe by Co generates a series compounds of $Li_2Co_8B_8$-$Li_4Co_8B_8$. We compute $E_{hull}$ for these structure and compared them to the existing Li-Co-B convex hull phase diagram in the Material Project and OQMD database [28,37]. As shown in Fig. 1(e), the $Li_x(CoB)_y$ phases also exhibit relatively low energies in the Li-Co-B system. Notably, we reveal that $Li_3Co_8B_8$ is also a thermodynamically stable phase which modifies the convex hull reported in [27]. This highlights the exceptional stability of the 3-8-8 structural framework across several members of the series of Li–transition metal borides.

## B. Lithium ion diffusion

The 1D channel structure in this $Li_x(FeB)_y$ series inspires us to study possible Li-ion diffusion. We carried out AIMD simulations for $Li_3Fe_8B_8$ at 300-600 K. The atomic trajectory at 300 K from the simulation is illustrated in Fig. 3(a). While the Fe and B atoms vibrate near their crystalline lattice sites, the Li atoms move rapidly along the 1D channels, behaving like a liquid. Such a characteristic suggests that the $Li_3Fe_8B_8$ compound is in a superionic state under ambient conditions. The MSD analysis yielded a room-temperature diffusion coefficient of $6.2 \times 10^{-4}$ cm$^2$/s, corresponding to an exceptionally high ionic conductivity of 65.3 S/cm. This remarkably high diffusion rate exceeds that of most superionic conductors [43–45]. Due to the high diffusion coefficient, extending the simulation time does not significantly alter the MSD (Supplementary Material Fig. S3 [36]). The Arrhenius fitting in Fig. 3(b) further revealed a low diffusion barrier of 0.02 eV.

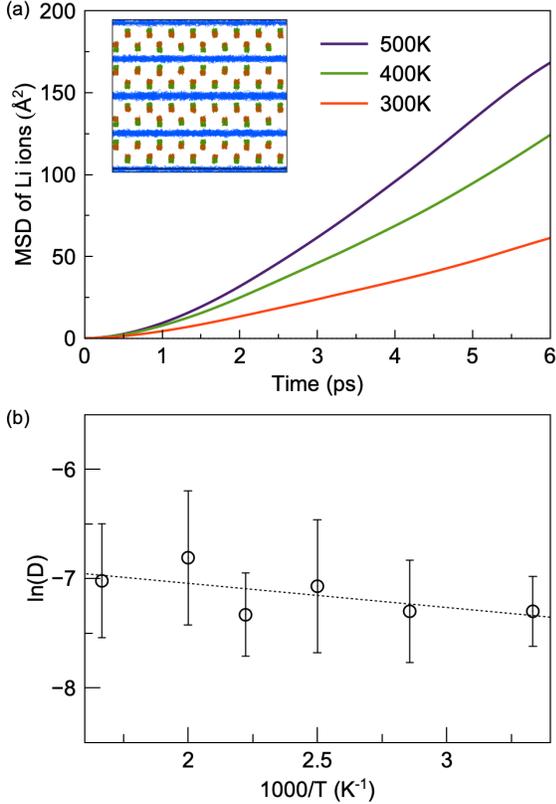

**FIG. 3** (a) Mean square displacement of Li ions in $Li_3Fe_8B_8$. The inset shows the AIMD trajectory of the $Li_{48}Fe_{128}B_{128}$ supercell at 300K. Blue is Li. Brown is Fe. Green is B. (b) Arrhenius plot of the Li ion diffusion coefficients obtained from MSD. Error bars represent the standard deviation from three independent simulations at each temperature. The Arrhenius fitting with $D = D_0 \exp\frac{-E_a}{k_B T}$ results in $E_a$=0.02 eV and $D_0 = 1.4 \times 10^{-3}$ cm$^2$/s.

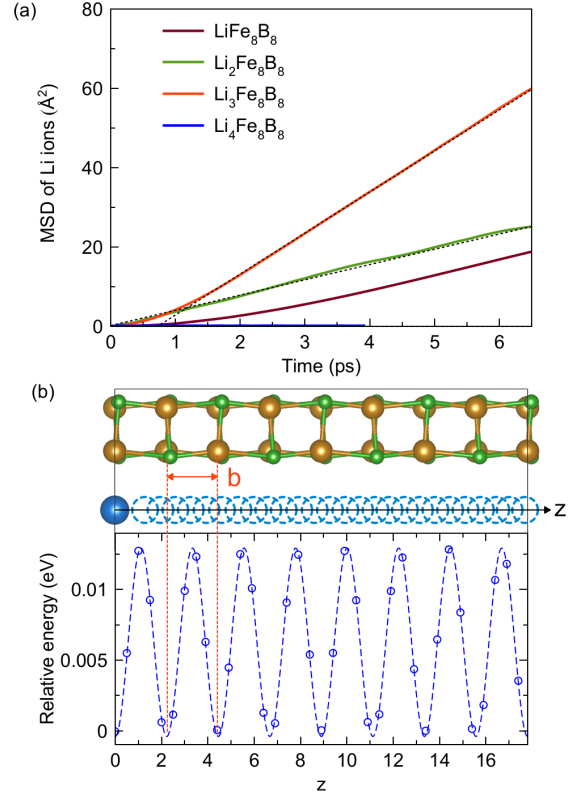

**FIG. 4** (a) Mean square displacement for Li ions in $Li_1Fe_8B_8$, $Li_2Fe_8B_8$, $Li_3Fe_8B_8$, $Li_4Fe_8B_8$. The dotted line is the linear fitting. (b) The periodic potential field along z direction caused by the $Fe_4B_4$ backbone structure. The upper panel illustrates a series of configurations with a single Li positioned at different sites in a $Fe_{32}B_{32}$ crystal structure. The lower panel shows the static energy as a function of Li ion positions. The dashed line is the fitting of a cosine function. $b$ is the period of the potential field.

The diffusion mechanism of Li ions in the 1D channels of $Li_3Fe_8B_8$ can be explained by the Frenkel–Kontorova model [46] which describes the motion of a particle chain within a periodic potential field. The equilibrium distance between neighboring atoms in the chain is denoted by $a$, and the period of the potential is denoted by $b$. When $a$ is an integer multiple of $b$, the system is in a *commensurate* state, with particle positions aligning with the periodic potential and becoming pinned at energy minima, which reduces the diffusion. Conversely, when $a$ is not an integer multiple of $b$, the system is *incommensurate*, where particles can move more freely, leading to high diffusion rates [47]. In the $Li_3Fe_8B_8$ structure, $a$ is the spacing between adjacent Li ions in the Li chain. The periodic potential field is caused by the $Fe_4B_4$ backbone structure along the direction of the Li chain. To identify the periodicity of the potential field, we remove the Li atoms from the $Li_3Fe_8B_8$ structure but manually position one single Li atom along the direction of the Li chain. By computing the static energy (without structure optimization) for configurations with different Li positions, we find the period $b$ of the potential field, corresponding to the distance between two energy minima in Fig. 4(b), is precisely the length of two adjacent $Fe_2B_2$ unit along the direction of the Li chain. Therefore, $Li_3Fe_8B_8$ is incommensurate, which may directly contribute to its high Li-atom diffusivity, as predicted by the FK model. To further confirm that the high diffusivity arises from the incommensurate state, we adjusted the Li content in $Li_3Fe_8B_8$ to construct commensurate structures: $Li_1Fe_8B_8$, $Li_2Fe_8B_8$, and $Li_4Fe_8B_8$. AIMD simulations were performed on each of these structures. The MSD curves of these systems, shown in Fig. 4(a), confirm that the incommensurate $Li_3Fe_8B_8$ exhibits the highest diffusion coefficient among the systems, consistent with the description from the Frenkel–Kontorova model.

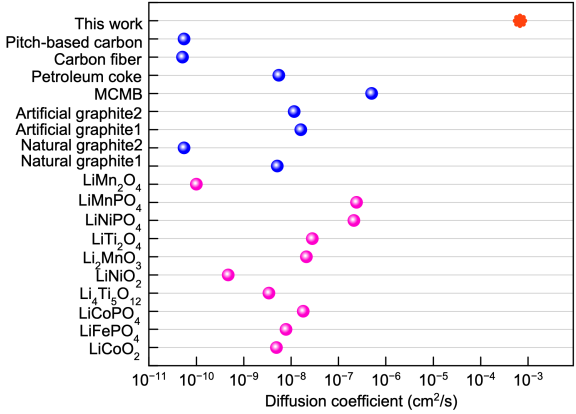

**FIG. 5** The diffusion coefficient of representative battery materials. Blue balls represent carbon-based conductive additive materials [48]. Pink represent common electrode materials [49].

### C. The application prospects of $Li_3Fe_8B_8$

Currently, commercial LiBs primarily use graphite as the anode material, which provides a theoretical specific capacity of 372 mA·h/g [50–52]. Silicon offers a much higher theoretical capacity of 4200 mA·h/g due to its ability to form a lithium-rich alloy, making it a promising alternative [53]. However, the application of Si anodes faces challenges, including the inherently low diffusion rates of Li ions and electrons which hinder the effectiveness of Si anodes in fast charging [54–58], and the substantial volume expansion during charge-discharge cycles which leads to particle fracturing and subsequent material loss [59]. Carbon-based materials, such as graphite [60,61], carbon black [61,62], graphene [63,64], and carbon nanotubes [65–67], are used as conductive additives to improve electronic conductivity and mechanical properties. However, limited lithium-ion diffusivity [48] in these conductive additives necessitates a search for alternative materials with improved ion transportation. Figure 5 shows that the diffusion coefficients of Li ions in both common electrode materials and carbon-based materials range from the order of $10^{-7}$ to $10^{-11}$ cm$^2$/s. In comparison, the $Li_3Fe_8B_8$ phase predicted in this work exhibits a diffusion coefficient of several orders of magnitude higher, indicating its excellent ion transport capability. Furthermore, the activation energy calculated from our AIMD simulations is only 0.02 eV, which is significantly lower than that of the well-known superionic conductor $Li_{10}GeP_2S_{12}$ (~0.25 eV) [68]. This exceptionally low migration barrier not only supports rapid ion diffusion but also implies weak temperature dependence, making $Li_3Fe_8B_8$ particularly attractive for battery applications operating under low temperatures or variable thermal environments without the need for extensive thermal management.

Table 1. Elastic constants ($C_{11}$, $C_{12}$, $C_{13}$, $C_{33}$, $C_{66}$), Young's modulus ($E$), bulk modulus ($B$), shear modulus ($G$) and Pugh's ratio ($B/G$) of tetragonal $Li_3Fe_8B_8$ phase.

| Elastic constants | Value (GPa) | Mechanical modulus | Value (GPa) |
|---|---|---|---|
| $C_{11}$ | 367.6 | $E$ | 267 |
| $C_{12}$ | 158.1 | $B$ | 206 |
| $C_{13}$ | 134.3 | $G$ | 104 |
| $C_{33}$ | 287.6 | $v$ | 0.284 |
| $C_{44}$ | 111.4 | $B/G$ | 1.99 |
| $C_{66}$ | 100.3 | | |

As a conductive additive for silicon anodes, it is also essential to exhibit decent mechanical properties to mitigate the volumetric expansion of the anode. Thus, we examine the mechanical properties of $Li_3Fe_8B_8$ by calculating its elastic constants in Table 1. According to the modified Born criterion [69], the tetragonal $Li_3Fe_8B_8$ phase satisfies the mechanical stability criteria. The bulk modulus, shear modulus, and Young's modulus are computed with the VRH approximation [70]. It is evident that the bulk modulus of $Li_3Fe_8B_8$ is significantly higher than the shear modulus, indicating that the compressive strength of this phase surpasses its shear strength, making it more resistant to compression. Poisson's ratio for the compound is 0.284, which exceeds the critical value of 0.26 [71], indicating that $Li_3Fe_8B_8$ is a ductile phase resistant to fracturing [72]. The Pugh criterion evaluates a material's ductility or brittleness by computing the ratio of its bulk modulus ($B$) to shear modulus ($G$). The $B/G$ ratio for the $Li_3Fe_8B_8$ phase is 1.99, which exceeds the critical value of 1.75, thereby confirming the ductile nature of $Li_3Fe_8B_8$ [72]. These data suggest the material's potential to form conductive networks without fracturing. Additionally, Young's modulus of $Li_3Fe_8B_8$ is 267 GPa, which is higher than that of steel (210 GPa) [73]. This indicates that $Li_3Fe_8B_8$ possesses strong resistance to deformation. Consequently, the conductive networks formed by this material can effectively mitigate the volume expansion of electrode materials.

We note that the use of $Li_3Fe_8B_8$ as a conductive additive can be likened to carbon nanotubes, with the key difference being that it represents an ordered arrangement of 1D channels, which offers distinct advantages. In contrast to the strong van der Waals forces and high aspect ratio of carbon nanotubes, which often cause entanglement and the formation of large bundles during their synthesis and application, the ordered 1D channel structure of $Li_3Fe_8B_8$ may mitigate these issues.

### D. Experimental synthesis

Four experimental attempts were made to synthesize $Li_3Fe_8B_8$ using a reactive lithium hydride powder precursor. Using alkali metal hydrides in place of the metal itself is advantageous owing to the improved mixing of the salt-like hydride with the other powder precursors, resulting in better diffusion than using chunks of the metal itself [15–17,20]. The utility of this unconventional synthesis method has been proven in several ternary boride systems containing alkali and alkaline-earth metals, such as Li-Ni-B and Mg-Co-B [15–17,20,74]. Use of a LiH precursor drastically reduced the annealing time required to produce $LiNi_3B_{1.8}$ compared to a direct reaction of the elements [15]. Uniform mixing of precursors enabled the discovery of $MgCo_3B_2$, the first reported compound in the Mg-Co-B system, where in situ X-ray diffraction experiment utilizing a $MgH_2$ precursor aided in pinpointing the optimal synthesis temperature [20]. In all cases, it was possible to obtain the ternary borides from direct reaction of the elements, but the reaction was faster and typically at lower temperature, yielding products of higher purity when a hydride precursor was used. To attempt $Li_3Fe_8B_8$ synthesis, powders of LiH, Fe, and B were combined in a 4:8:8 molar ratio, ball-milled, and annealed at four different temperatures of 1023 K, 1073 K, 1173 K and 1273 K with variable dwelling time (samples A-D). Typically, a slight excess of the alkali precursor is required to form a target phase with a hydride reaction. PXRD performed on the synthesized powders in Fig. 6 gave no indication of a ternary phase; rather, iron-boron binary compounds were formed, with the unreacted lithium present post-synthesis presumably reacting with air and moisture to form amorphous or weakly-diffracting oxide or hydroxides. FeB is known to exist in two polymorphs. $\alpha$-FeB tends to form as a stochastic stacking intergrowth of two structures: CrB-type (space group $Cmcm$, abbreviated $oS$-FeB in Fig. 6) and $\beta$-FeB-type (space group $Pnma$, abbreviated $oP$-FeB in Fig. 6), with the relative abundance of $\beta$-FeB-type layers increasing with temperature [75]. On the other hand, it is known that $\beta$-FeB lacks the stacking faults of the $\alpha$ polymorph and exists as a long-range-ordered $P$-orthorhombic structure. This behavior aligns with the collected PXRD patterns: at lower temperatures (samples A and B), the profile more closely resembles $oS$-FeB, while at higher temperatures (samples C and D), peaks corresponding to $oP$-FeB became more prominent. Additionally, the phase fraction of $Fe_2B$ is observed to increase with temperature, although the phase is present in all samples. No unidentified peaks were found in any of the samples.

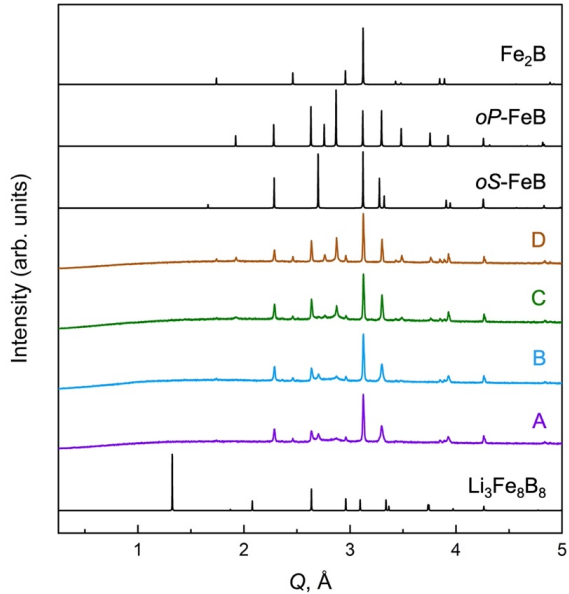

**FIG 6.** Experimental PXRD patterns for synthesis attempts (samples A–D) compared to calculated diffraction profiles for predicted $Li_3Fe_8B_8$, and experimentally observed $Fe_2B$, and two polymorphs of FeB.

Of the four heating profiles attempted in an effort to synthesize $Li_3Fe_8B_8$ from lithium hydride, iron, and boron precursors, none was successful in producing the desired ternary phase. The lack of evidence of any reaction between Li and Fe or B likely indicates that appropriate synthesis conditions have not yet been found to induce a reaction between volatile and reactive Li and high-melting Fe and B, implying that the $Li_3Fe_8B_8$ phase, if it exists, is probably synthesizable only over a narrow temperature range. In such a case, in situ PXRD on unreacted LiH, Fe, and B precursors loaded and sealed in a silica capillary would shed light on the appropriate temperature range for targeted synthesis of the compound, as was found to be the case with Mg-containing ternary boride systems [20,26]. In particular, in situ PXRD was used to pinpoint an appropriate temperature range for synthesis of $MgCo_3B_2$, a compound that was found to be sensitive to small variations in synthesis temperature and loading composition [20]. Applied pressure during the synthesis, e.g. high-pressure-high-temperature synthesis, is another synthetic approach that could be exploited to promote formation of the target ternary phase, in conjunction with the enhanced reaction kinetics provided by the use of hydride precursors.

## IV. Conclusion

In summary, we performed crystal structure prediction for ternary Li-Fe-B compounds and identified a thermodynamically stable $Li_3Fe_8B_8$ phase. The crystal structure of $Li_3Fe_8B_8$ features 1D channels that facilitate rapid lithium-ion transport. *Ab initio* molecular dynamics simulations revealed that $Li_3Fe_8B_8$ exhibits a significantly higher Li-ion diffusion rate than common electrode and conductive additive materials by several orders of magnitude. The Frenkel–Kontorova model explains the enhanced diffusion in $Li_3Fe_8B_8$, which arises from the incommensurate state between the Li diffusion chain and the FeB backbone structures. The favorable mechanical properties of $Li_3Fe_8B_8$ suggest its ability to form conductive networks, which can be used to address common problems such as electrode degradation and volume expansion in silicon anodes. We also identified $Li_3Fe_8B_8$ as an itinerant ferromagnet near the magnetic quantum critical point. Such ferromagnetism makes this system sensitive to the external magnetic field, which can be used during manufacturing, recycling and safety monitoring of corresponding lithium-ion batteries. Several experimental attempts at synthesis of $Li_3Fe_8B_8$ using the hydride route did not realize the phase, underlining the difficulty of achieving a reaction between elements with drastically different reactivities and melting points. Synthesis of $Li_3Fe_8B_8$ likely depends on careful optimization of several parameters and could be aided by the use of in situ synchrotron PXRD. Altogether, the 1D channel, along with the ferromagnetism, structural and mechanical stability of $Li_3Fe_8B_8$, underscores its potential in battery technology. Further experimental efforts to validate these predictions are highly desirable.

**ACKNOWLEDGMENTS**
We are grateful to Xingke Cai and Zonghua Pu for insightful discussions. The work at Xiamen University was supported by the National Natural Science Foundation of China (Grant No. T2422016), the Natural Science Foundation of Xiamen (Grant No. 3502Z202371007), and the Fundamental Research Funds for the Central Universities (Grant No. 20720230014). P.O., Z.Z., J.V.Z. acknowledge financial support from the U.S. Department of Energy (DOE) Established Program to Stimulate Competitive Research (EPSCoR) Grant No. DE-SC0024284. V.A. was supported by the U.S. Department of Energy, Office of Basic Energy Sciences, Division of Materials Sciences and Engineering. Ames National Laboratory is operated for the U.S. Department of Energy by Iowa State University under Contract No. DE-AC02-07CH11358.

Lithium−Oxygen Battery, Adv. Mater. **36**, 2312661 (2024).

# Prediction of $Li_3Fe_8B_8$ compound with rapid one-dimensional ion diffusion channels

Shiya Chen[1], Paul Oftedahl[2], Zhen Zhang[3], Zepeng Wu[1], Junjie Jiang[1], Vladimir Antropov[4], Julia V. Zaikina,[2] Shunqing Wu[1], Kai-Ming Ho[3], Yang Sun[1]

[1]*Department of Physics, Xiamen University, Xiamen 361005, China*
[2]*Department of Chemistry, Iowa State University, Ames, Iowa 50011, United States*
[3]*Department of Physics, Iowa State University, Ames, IA 50011, United States*
[4]*Ames National Laboratory, U.S. Department of Energy, Ames, IA 50011, United States*

**Contents of this file**
  Figure S1-S3
  Table S1-S2

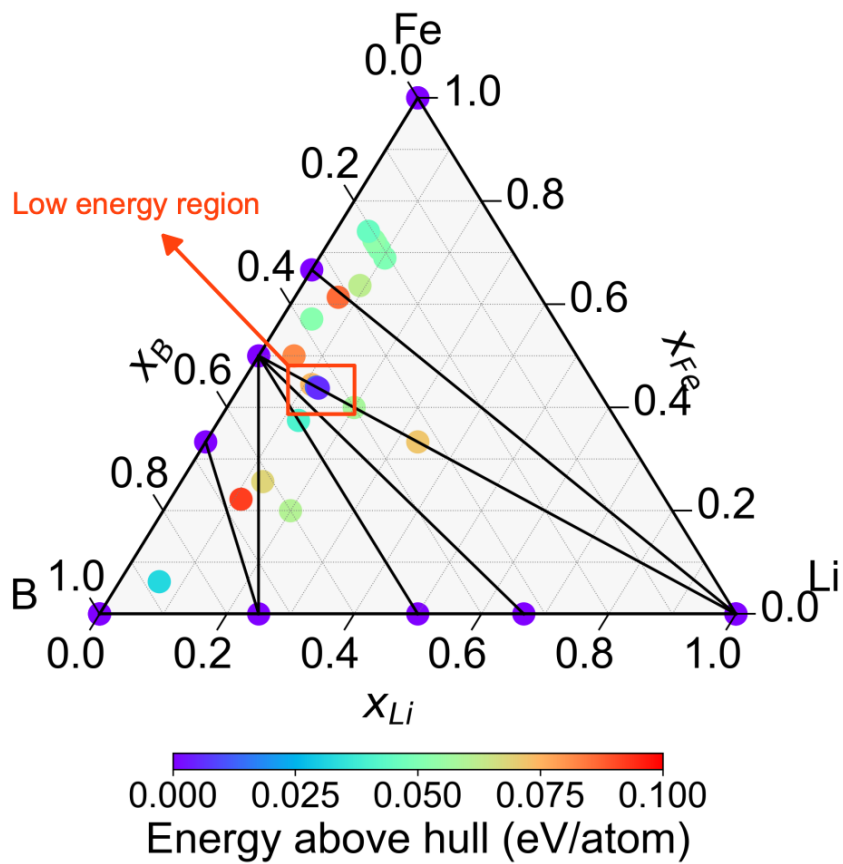

**Fig. S1** The Li-Fe-B convex hull with substituted structures. The red box highlights a series of low-energy metastable phases represented by $Li_x(FeB)_y$.

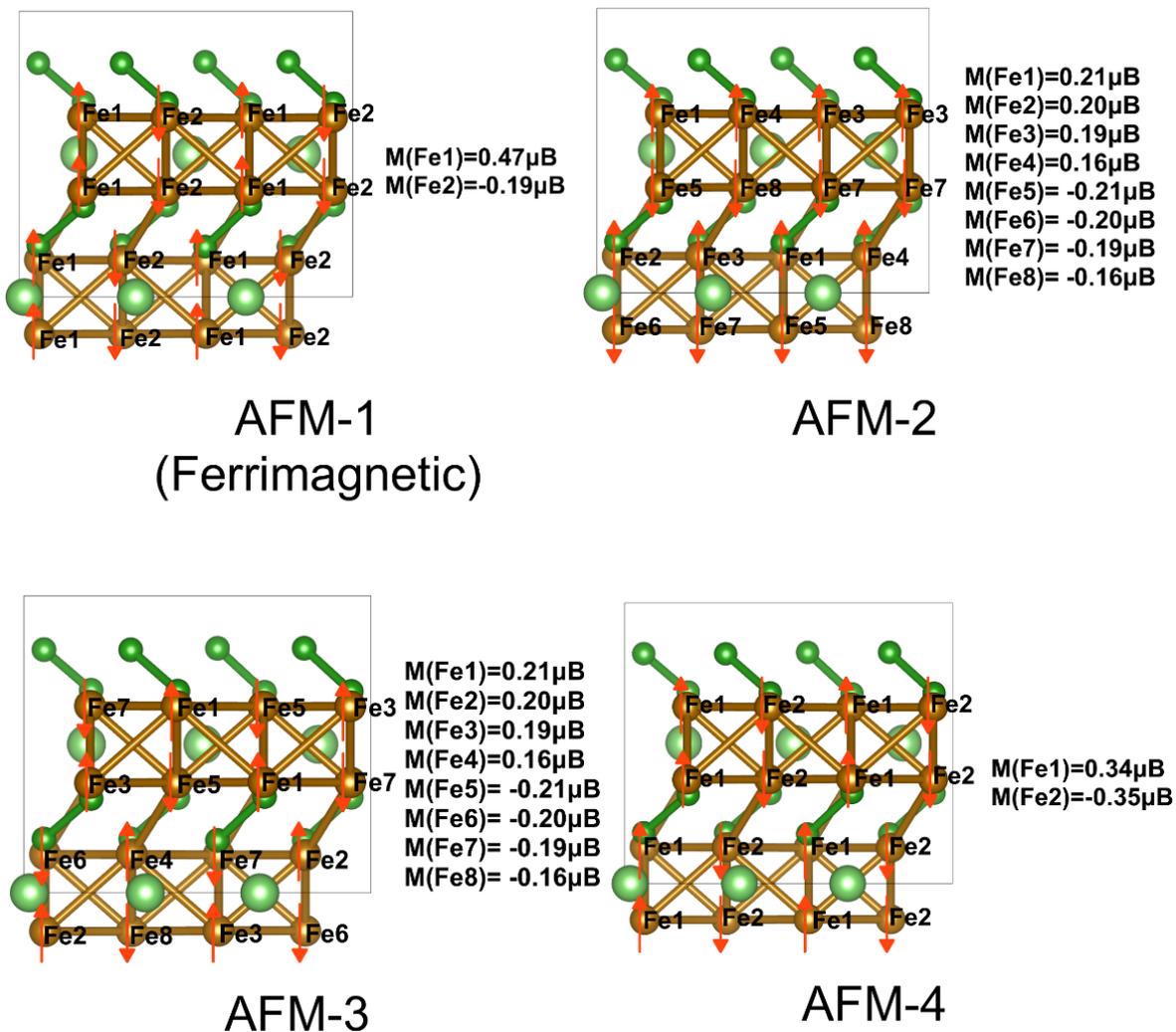

**Fig. S2** Different magnetic configurations of $Li_3Fe_8B_8$. Arrows indicate the directions of magnetic moments for each atom. The energy of these configurations are summarized in Table S2.

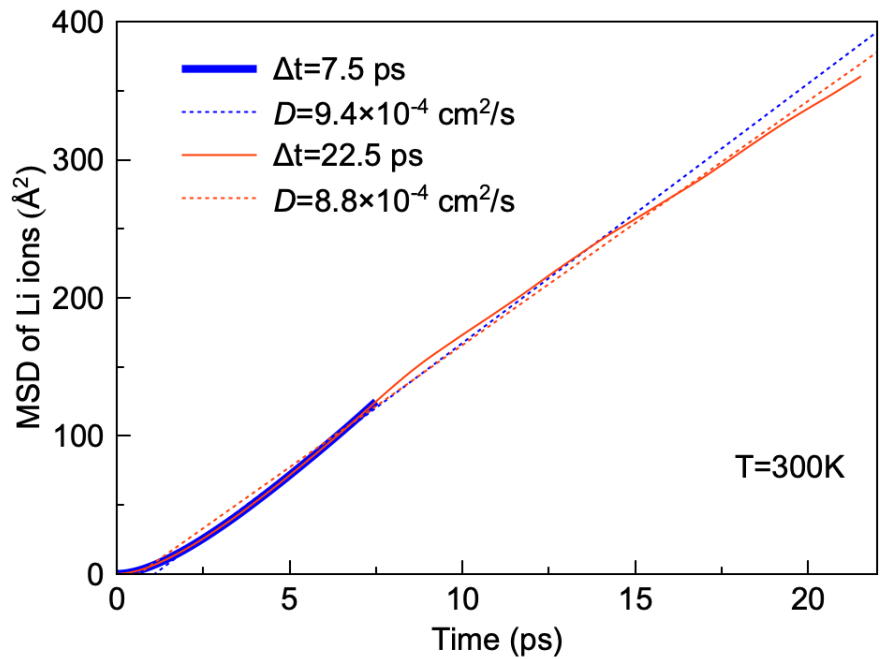

**Fig. S3** Mean-squared displacement (MSD) of Li ions in $Li_3Fe_8B_8$ at 300 K. The diffusion coefficients ($D$) computed using the trajectories of 7.5 ps and 22.5 ps are consistent.

**Table S1** Crystallographic data of ternary $Li_3Fe_8B_8$ phases.

| Phase | Space group | Lattice param | Wyckoff site | Atom | x | y | z |
|---|---|---|---|---|---|---|---|
| $Li_3Fe_8B_8$ | $P\text{-}421c$ | a = 6.711 | 2a | Li1 | 0.000 | 0.000 | 0.000 |
| | | b = 6.711 | 4c | Li2 | 0.000 | 0.000 | 0.335 |
| | | c = 7.838 | 8e | B1 | 0.180 | 0.677 | 0.057 |
| | | $\alpha = \beta = \gamma = 90°$ | 8e | B2 | 0.182 | 0.324 | 0.307 |
| | | | 8e | Fe1 | 0.629 | 0.128 | 0.440 |
| | | | 8e | Fe2 | 0.630 | 0.873 | 0.190 |

**Table S2** Energies and magnetic moments of different magnetic configurations of $Li_3Fe_8B_8$.

| Configuration | $E_0$ (eV/$Li_6Fe_{16}B_{16}$) | $\Delta E$ (eV/Fe) | Tot moment ($\mu_B/Fe_{16}$) |
|---|---|---|---|
| FM | -262.886 | 0 | 12.52 |
| NM | -261.969 | 0.057 | 0.00 |
| AFM-1 | -262.134 | 0.047 | 2.04 |
| AFM-2 | -261.981 | 0.057 | 0.00 |
| AFM-3 | -261.802 | 0.068 | 0.00 |
| AFM-4 | -262.070 | 0.051 | 0.07 |